\documentclass[usenatbib,useAMS]{mn2e}
\usepackage{epsfig}
\usepackage{times}
\usepackage{epsfig}
\usepackage{graphicx}
\usepackage{bm}
\usepackage{dcolumn}
\usepackage{graphicx}
\usepackage{epstopdf}
\usepackage{color}
\usepackage{epsf}
\usepackage{epsfig}
\usepackage{graphicx,epic,eepic,color}
\graphicspath{{fig/}}

\begin{document}
\title[Photometric, Astrometric and Polarimetric observations of microlensing]{Photometric, Astrometric and Polarimetric observations of gravitational microlensing events}
\author[Sedighe Sajadian and Sohrab Rahvar]
{Sedighe Sajadian$^{1}$ and Sohrab Rahvar$^{2}$ \\ %\thanks{srahvar@pitp.ca}  \\
$^1$ School of Astronomy, Institute for Research in Fundamental Sciences (IPM), P.O. Box 19395-5531, Tehran, Iran \\
$^2$ Department of Physics, Sharif University of Technology, P.O. Box 11155-9161, Tehran, Iran \\
}

\maketitle
\begin{abstract}
The gravitational microlensing as a unique astrophysical tool can be
used for studying the atmosphere of stars thousands of parsec far
from us. This capability results from the bending of light rays in
the gravitational field of a lens which can magnify the light of a
background source star during the lensing. Moreover, one of
properties of this light bending is that the circular symmetry of
the source is broken by producing distorted images at either side of
the lens position. This property makes the possibility of the
observation of the polarization and the light centroid shift of
images. Assigning vectors for these two parameters, they are
perpendicular to each other in the simple and binary microlensing
events, except in the fold singularities. In this work, we
investigate the advantages of polarimetric and astrometric
observations during microlensing events for (i) studying the surface
of the source star and spots on it and (ii) determining the
trajectory of source stars with respect to the lens. Finally we
analyze the largest sample of microlensing events from the OGLE
catalog and show that for almost $\sim 4.3\%$ of events in the
direction of the Galactic bulge, the polarization signals with large
telescopes would be observable.

\end{abstract}
%%%%%%%%%%%%%%%%%%%%%%%%%%%%%%%%%%%%%%%%%%%%%%%%%%%%%%%%%%%%

\section{Introduction}
% ---------------- introducing microlensing ------------------------------------------
The gravitational field of an astrophysical object can deviate the
light path of a background source star and the result is the
formation of two images at either side of the lens
\cite{Einstein36}. In the Galactic scale, the angular separation of
the images is of the order of milli arc second, too small to be
resolved by the ground-based telescopes. Instead, the overall light
from the images received by the observer is magnified in comparison
to an un-lensed source. This phenomenon is called gravitational
microlensing and has been proposed as an astrophysical tool to probe
dark objects in the Galactic disk. In recent years it has been used
also for discovering the extra solar planets, the stellar atmosphere
of distant stars and etc. \cite{Leibes,ChangRefesdal,Paczynski86}. A
detailed review on these topics can be found in Gaudi (2012). An
important problem in photometric observations of microlensing events
is the degeneracy between the lens parameters as the distance, mass
and velocity. We can partially resolve this degeneracy by using
higher order terms as the parallax and finite size effects which can
slightly change simple microlensing light curves. However, these
effects can not be applied for all the microlensing events. There
are other observables such as (i) the centroid shift of images and
(ii) the polarization variation of images during the lensing that
can break this degeneracy. Here, we assume that during photometric
observations of microlensing events, polarimetric and astrometric
observations also can be done.

% ------------------------ introducing astrometric centroid shift --------------------
{\bf Astrometric observation}: In the gravitational microlensing,
the light centroid of images deviates from the position of the
source star and for the case of a point-mass lens, the centroid of
images traces an elliptical shape during the lensing
\cite{Walker,Miyamoto,Hog,Jeong}. By measuring the light centroid
shift of images with the high resolution ground-based or space-based
telescopes, accompanied by measurements of the parallax effect, the
lens mass can be identified \cite{Paczynski96,Miralda96}.
%In this case we do not need to know
%exactly the distance of the lens and the source from the observer.
 This method is also applicable for
studying the structure of the Milky way with enough number of
microlensing events \cite{Rahvar2005}. Also the degeneracy in the
close and the wide caustic-crossing binary microlensing events can
be removed by astrometric observations
\cite{Dominik,Gould,Chung}. \\
%Here we will assume zero order astrometric deviation
%without taking into account the higher order terms as the parallax effect.

% ------------------- introducing polarization effect ---------------------------------
{\bf Polarimetric observation}: Another feature in the gravitational
microlensing is the time-dependent variation of the polarization
during the lensing. The scattering of photons by electrons in the
atmosphere of star makes a local polarization in different positions
over the surface of a star and as a result of circular symmetry of
the star, the total polarization is zero \cite{chandrasekhar60,sob}.
During the gravitational microlensing, the circular symmetry of
images breaks and the total polarization of source star is non-zero
and changes with time \cite{schneider87,simmons95a,Bogdanov96}.
Measuring polarization during microlensing events helps us to
evaluate the finite source effect, the Einstein radius and the
limb-darkening parameters of the source star
\cite{yoshida06,Agol96,schneider87}.

% ----------------------------- OUR PROPOSAL AND ITS RESULTS ------------------------
In this work we start with the investigation for a possible relation
between the polarization and the centroid shift vectors. We find an
orthogonality relation between them in the simple as well as the
binary lensing. However, near the cusp in the caustic lines of a
binary lens, the polarization and centroid shift vectors are not
normal to each other except on the symmetric axis of the cusp. This
effect enables us to discover the source trajectory relative to the
caustic line. As a result of this orthogonality relation, the
polarimetry measurements can resolve the source trajectory
degeneracy, i.e. $u_{0}$ degeneracy, in the same way as the
astrometric observations. Finally, we study the effects of source
spots as a perturbation effect and show that they can break this
orthogonality relation. Studying the time variation of polarization
can provide a unique tool to distinguish and study the source
anomalies such as spots on the surface of the source stars.

The layout of the paper is as follows. In sections (\ref{ortho}) we
introduce the polarization and astrometry in the gravitational
lensing and demonstrate that there is an orthogonality relation
between them in simple microlensing events. We extend this
discussion for the binary lensing in section (\ref{twolens}) and
discuss how this correlation helps to resolve the degeneracy
problem. In section (\ref{spot}), we investigate the effect of
source spots in the polarimetric observation. Finally, we take the
largest sample of the OGLE microlensing data and estimate the number
of events with the observable polarimetry signals. We conclude in
section (\ref{result}).

%%%%%%%%%%%%%%%%%%%%%%%%%%%%%%%%%%%%%%%%%%%%%%%%%%%%%%%%%%%%%%%%%%%%%%%%%%%%%%%%%%%%%%

\section{Polarimetric and Astrometric shift in gravitational microlensing}
\label{ortho} In this section we first review the polarization
during microlensing events. Then we study the astrometric shift and its
orthogonality with the polarization for an extended source star.

\subsection{Polarization during gravitational microlensing}
Chandrasekhar (1960) has shown that photons can be scattered by
electrons (Thompson scattering) in the atmosphere of hot stars which
makes a linear and local polarization. The amount of polarization
enhances from the centre to the limb and at a given wavelength in
each point it is proportional to the cosine of the angle between the
line of sight and the normal vector to the star surface. The other
mechanisms such as photon scattering on atomic, molecular species
and neutral hydrogen (Rayleigh scattering) or on dust grains are
also responsible for producing a local polarization over late-type
main sequence and cool giant stars \cite{Ingrosso}. Due to the
circular symmetry of the star surface, the total light of a distant
star is unpolarized. This circular symmetry can be broken by spots
on the star surface, magnetic fields or the lensing effect and as a
result, we expect to detect a non-zero polarization for these cases.
For example, the emission lines from T-Tauri stars (pre-main
sequence stars) show a linear polarization due to the light
scattering by dust grains in the circumstellar disk around the
central stars. This polarization changes with time due to variations
in the configuration of the dust pattern
\cite{Drissen1989,Akitaya2009}.

Schneider \& Wagoner (1987) noticed that there is a net and
time-dependent polarization for a lensed supernova. They estimated
the amounts of polarization degree near point and critical line
singularities. The existence of a net polarization during
microlensing events due to circular symmetry breaking was noticed by
Simmons et al. (1995a,b) and Bogdanov et al. (1996). Also,
polarization in binary microlensing events was numerically
calculated by Agol (1996). He noticed that in a binary microlensing
event, the net polarization is larger than that by a single lens and
reaches to one percent during the caustic crossing. A
semi-analytical formula for polarization degree induced by a single
microlens was derived by Yoshida (2006). Recently, Ingrosso et al.
(2012) evaluated the expected polarization signals for a set of
reported high-magnification single-lens and exo-planetary
microlensing events towards the Galactic bulge. They showed that it
reaches to $0.04$ per cent for late-type stars and rises to a few
per cent for cool giants.

\begin{figure}
\begin{center}
\psfig{file=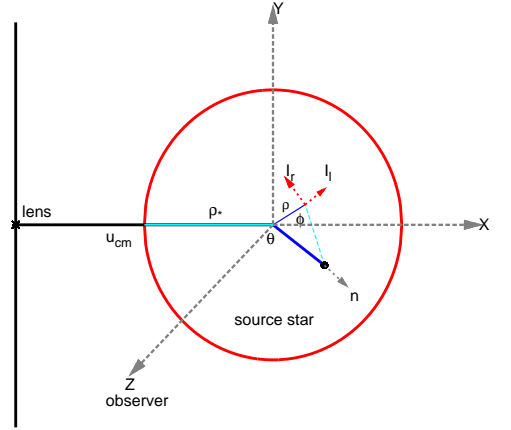,angle=270,width=11.cm,clip=0} \caption{
Demonstration of a projected source surface on the lens plane (red
circle). In this figure, the black star and the black spot represent the
lens position and a typical point on the source surface,
respectively. Source star is projected on the sky plane. The
directions of $\boldsymbol{n}$ and $\boldsymbol{Z}$ refer to the
propagation direction and line of sight towards the observer.
$u_{cm}$ connects lens position to the source center, $\phi$ is
the azimuthal angle between the lens-source connection line and the
line from the origin to each projected element over the source
surface and $\theta$ is the projection angle i.e. $\rho=\sin
\theta$. $\boldsymbol{l_l}$ and $\boldsymbol{l_r}$ are two unit
vectors normal to the direction of
$\boldsymbol{Z}$ as shown in the figure.}\label{coordinate}
\end{center}
\end{figure}

In this part, we review how to calculate the net polarization of a
source star in the microlensing events. For describing the polarized
light, we use $S_{I}$, $S_{Q}$, $S_{U}$ and $S_{V}$ as the Stokes
parameters. These parameters show the total intensity, two
components of linear polarizations and the circular polarization
over the source surface, respectively \cite{Tinbergen96}. For a
stellar atmosphere, we set the circular polarization to be zero,
$S_{V}=0$. The linear polarization degree $(P)$ and angle of
polarization $(\theta_{p})$ as functions of total Stokes parameters
are given by \cite{chandrasekhar60}:
\begin{eqnarray}
P&=&\frac{\sqrt{S_{Q}^{2}+S_{U}^{2}}}{S_{I}},\nonumber\\
\theta_{p}&=&\frac{1}{2}\tan^{-1}{\frac{S_{U}}{S_{Q}}},
\end{eqnarray}
where the un-normalized Stokes parameters are defined as follows:
\begin{eqnarray}
S_{Q}&\equiv&\int_{\mathcal{S}}d^{2}\mathcal{S}<E_{X}E_{X}-E_{Y}E_{Y}>= \nonumber\\
&=&-\int d^{2}\mathcal{S}~I_{-}(\mu)\cos(2\phi),\nonumber\\
S_{U}&\equiv&-\int_{\mathcal{S}}d^{2}\mathcal{S}<E_{X}E_{Y}+E_{Y}E_{X}>=\nonumber\\
&=&\int d^{2}\mathcal{S}~I_{-}(\mu)\sin(2\phi),\nonumber\\
S_{I}&\equiv&\int_{\mathcal{S}}d^{2}\mathcal{S}<E_{X}E_{X}+E_{Y}E_{Y}>=\int
d^{2}\mathcal{S}~I(\mu),
\end{eqnarray}
where $\mathcal{S}$ refers to source area projected on the lens
plane, $\phi$ is the azimuthal angle between the lens-source
connection line and the line from the origin to each element over
the source surface,  $I(\mu)=I_{l}(\mu)+I_{r}(\mu)$,
$I_{-}(\mu)=I_{r}(\mu)-I_{l}(\mu)$ and $<>$ refers to the time
averaging (see Figure \ref{coordinate}). In following, we adapt the
coordinate sets used by Chandrasekhar. Let us define
$\boldsymbol{n}$ being the normal to the source surface at each
point which is the propagation direction and $\boldsymbol{Z}$ being
the direction towards the observer. We define $\boldsymbol{r}$ and
$\boldsymbol{l}$ so that $I_{l}(\mu)$ being the emitted intensity by
the star atmosphere in the plane containing the line of sight and
the normal to the source surface on that point, i.e.
$(\boldsymbol{n})$, and $I_{r}(\mu)$ being the emitted intensity in
the normal direction to that plane, where $\boldsymbol{r} \times
\boldsymbol{l}= \boldsymbol{Z}$, $\mu=\cos(\theta)=\sqrt{1-
\rho^{2}}$ and $\rho$ is the distance from the centre to each
projected element over the source surface normalized to the
projected radius of star on the lens plane. Indeed, $\boldsymbol{l}$
represents the radial and $\boldsymbol{r}$ is the tangent
coordinates perpendicular to the line of sight. We consider a fixed
cartesian coordinate at the star centre where $\boldsymbol{X}$-axis
is outwards the lens position, $\boldsymbol{Y}$-axis is normal to
it, projected in the sky plane and $\boldsymbol{Z}$-axis is in the
line of sight towards the observer. The projected source surface on
the lens plane (red circle) and the specified axes are shown in
Figure (\ref{coordinate}).

\begin{figure}
\begin{center}
\psfig{file=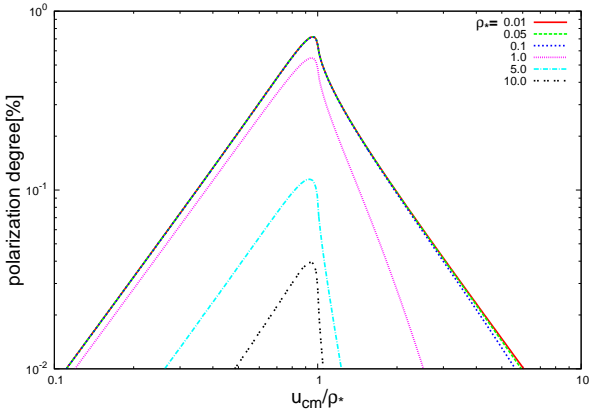,angle=270,width=8.cm,clip=} \caption{The
polarization degree versus $u_{cm}/\rho_{\star}$ for different
values of star size and impact parameter. The polarization raises to
a maximum value at $u_{cm}/\rho_{\star}\approx 0.96$
( Schneider \& Wagoner 1987).} \label{figpol}
\end{center}
\end{figure}

In simple microlensing events, the Stokes parameters by
integrating over the source surface are given by:
\begin{eqnarray}\label{tsparam}
\left( \begin{array}{c} S_{Q}\\
S_{U}\end{array}\right)&=&\rho^2_{\star}\int_{0}^1\rho~d\rho\int_{-\pi}^{\pi}d\phi I_{-}(\mu) A(u) \left( \begin{array}{c} -\cos 2\phi \\
\sin 2\phi \end{array} \right),\nonumber\\
S_{I}&=&~\rho^2_{\star}\int_{0}^{1}\rho d\rho\int_{-\pi}^{\pi}d\phi
I(\mu)~ A(u),
\end{eqnarray}
where $\rho_\star$ is the projected radius of star on the lens plane
and normalized to the Einstein radius, $u=(u_{cm}^2+ \rho^2
\rho^{2}_{\star}+2 \rho \rho_{\star} u_{cm} \cos\phi)^{1/2}$ is the
distance of each projected element over the source surface with
respect to the lens position, $u_{cm}$ is the impact parameter of
the source centre and magnification factor for a simple microlensing
is $$A(u)=\frac{u^2+2}{u \sqrt{u^2+4}}.$$ The amounts of $I(\mu)$
and $I_{-}(\mu)$ by assuming the electron scattering in spherically
isotropic atmosphere of an early-type star were evaluated by
Chandrasekhar (1960) as follows
\begin{eqnarray}
I(\mu)&=&I_{0}(1-c_{1}(1-\mu)),\nonumber\\
I_{-}(\mu)&=&I_{0}c_{2}(1-\mu),
\end{eqnarray}
where $c_{1}=0.64$, $c_{2}=0.032$ and $I_{0}$ is the total intensity
emitted towards the line of sight direction \cite{schneider87}.

For a point-mass lens, the integrals over the azimuthal angle $\phi$
of the total Stokes parameters, i.e. equation (\ref{tsparam}), are
reduced to a combination of complete elliptical integrals
\cite{yoshida06}. We calculate the Stokes parameters by numerical
integrations. Figure (\ref{figpol}) represents the dependence of the
polarization degree on the source size, $\rho_{\star}$ and
$u_{cm}/\rho_{\star}$. The polarization has its maximum value at
$u_{cm}/\rho_{\star}\approx0.96$ \cite{schneider87}. If
$u_{0}<\rho_{\star}$ there are two times in which $u_{cm}=0.96
\rho_{\star}$, so the time profile of polarization has two peaks
(transit case) whereas if $u_{0}\geq \rho_{\star}$ only in the
closest approach the time profile of the polarization has a peak
(bypass case) \cite{simmonsb}. For the case that the finite size
effect of source star is small, the probability of detecting
polarization also is small as the chance that the lens and source
approach with the impact parameter comparable to the source size is
small. Hence, detecting polarization effect is in favour of
microlensing events with a large finite size parameter (i.e. giant
stars).

According to equation (\ref{tsparam}), the total Stokes parameter
$S_{U}$ for a limb-darkened and circular source in simple
Paczy\'nski microlensing events is zero and $S_{Q}$ is negative.
Hence, the net polarization is normal to the lens-source connection
line. In figure (\ref{figm}), we show the polarization map around a
point-mass lens located at the centre of the plane with solid black
lines. Note that, the size of lines is proportional to the
polarization. The orientation (i.e. $\theta_{p}$) is given in terms
of its angle with respect to the x-axis specified in Figure
(\ref{coordinate}). The arrow sign represents the centroid shift of
images that will be discussed in the next part.

\subsection{Astrometric centroid shift}
\begin{figure}
\begin{center}
\psfig{file=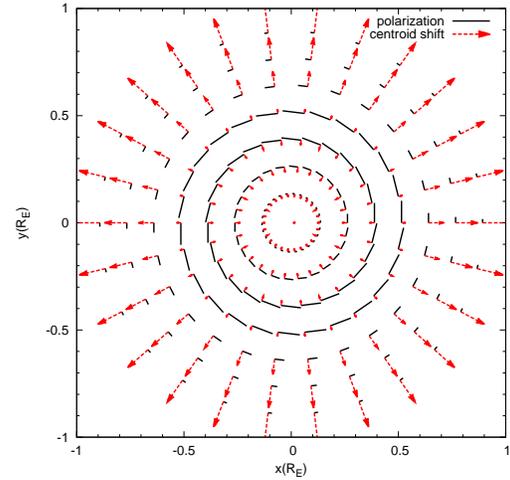,angle=270,width=9.cm,clip=0} \caption{The
astrometric (red dashed vectors) and polarimetric maps (black solid
lines) around a point-mass lens located at centre of plane. We set
$\rho_{\star}=0.5$. Note that, the size of centroid shifts and
polarizations are normalized by the arbitrary factors. For
$u\simeq\rho_{\star}$, the size of the centroid shift vector tend
to zero while polarization signal is maximum.} \label{figm}
\end{center}
\end{figure}

In gravitational microlensing events a shift in the light
centroid of the source star happens with respect to its real
position. For a point-mass lens the
astrometric shift in the light centroid is defined by:
\begin{eqnarray}
\boldsymbol{\delta\theta}_{c}=\frac{\mu_{1}\boldsymbol{\theta_{1}}+
\mu_{2}\boldsymbol{\theta_{2}}}{\mu_{1}+\mu_{2}}-\boldsymbol{u}\theta_{E}
=\frac{\theta_{E}}{u^{2}+2}\boldsymbol{u},
\end{eqnarray}
where $\theta_{E}$ is the angular size of the Einstein ring. The
astrometric shift traces an ellipse, so-called the \emph{astrometric
ellipse} while the source star passes an straight line with respect
to the lens plane \cite{Walker,Jeong}. The ratio of axes for this
ellipse is a function of the impact parameter and for large impact
parameters this ellipse converts to a circle whose radius decreases
by increasing the impact parameter and for small values, it turns to
a straight line \cite{Walker}. In Figure (\ref{figm}), we plot a
vector map containing the normalized astrometric centroid shift at
each position of the lens plane which is a symmetric map around the
position of a point-mass lens. All the vectors in this case are
radial and outward.

\begin{figure}
\begin{center}
\psfig{file=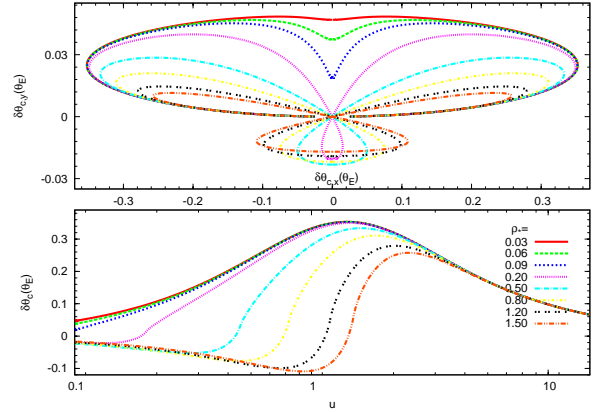,angle=270,width=8.cm,clip=} \caption{The
normalized centroid shift trajectories in the lens plane (upper panel) and
the normalized centroid shift amount versus $u$ (lower panel) for
different amounts of $\rho_{\star}$ and  $u_{0}=0.1$. The negative
value of centroid shift (in lower panel) indicates reverting its
direction with respect to the $x$-axis which occurs for $u\leq
\rho_{\star}$.} \label{cenfin}
\end{center}
\end{figure}

The astrometric centroid of an extended source is given by (Witt \&
Moa 1998)
\begin{eqnarray}\label{astro}
\boldsymbol{\theta}^{fs}_{c}=\frac{1}{\pi \rho^{2}_{\star}
\bar{I}\mu^{fs}}\sum_{i}\int_{\mathcal{S}}d^{2}\mathcal{S}~
I(\boldsymbol{u}) \boldsymbol{\theta}_{i}\mu_{i},
\end{eqnarray}
where $\bar{I}$ is the mean surface brightness of source,
$\boldsymbol{\theta}_{i}$ is the position of the $i$th image with
the magnification factor $\mu_{i}$ and $\mu^{fs}$ is the total
magnification factor of an extended source:
\begin{eqnarray}
\mu^{fs}=\frac{1}{\pi \rho^{2}_{\star}
\bar{I}}\sum_{i} \int_{\mathcal{S}}d^{2}\mathcal{S}~I(\boldsymbol{u})\mu_{i}(\boldsymbol{u}).
\end{eqnarray}
In a simple microlensing event, where we have a single lens and the
surface brightness of source star is uniform,
$\sum_{i}\boldsymbol{\theta}_{i}\mu_{i}=\theta_{E}\frac{u^2+3}{
\sqrt{u^2+4}} \hat{u}$ and the normalized centroid shift for an
extended source is given by:
\begin{eqnarray}\label{astrop}
\boldsymbol{\delta\theta}_{c}^{fs}(\theta_{E})=-\boldsymbol{u_{cm}}+\frac{1}{\pi \mu^{fs}}\int_0^{1} \rho d\rho \int_{-\pi}^{\pi} d\phi I(\rho) \frac{u^2+3}{\sqrt{u^2+4}} \hat{u}, %\left( \begin{array}{c} \rho \cos\phi+ u_{cm} \\
%\rho \sin\phi \end{array} \right).
\end{eqnarray}
where $\hat{u}=(\rho \cos\phi+ u_{cm} ,\rho \sin\phi)/u$ (see Figure
\ref{coordinate}). In Figure (\ref{cenfin}), we plot the
trajectories of normalized centroid shifts in the lens plane (upper
panel) and the absolute value of normalized centroid shifts versus
$u$ (lower panel) for different values of $\rho_{\star}$ and fixed
value of $u_{0}=0.1$.

In the case of a uniform or limb-darkened source, the centroid shift
component normal to the lens-source connection line (i.e. $Y$-axis
in Figure \ref{coordinate}) is zero. On the other hand, the
polarization orientation is normal to the centroid shift for a
single lens (see Figure \ref{figm}). The anomalies over the source
surface can break this orthogonal relation. We note that,
astrometric signals are detectable for the nearby and massive lenses
which have large angular Einstein radii. On the other hand
polarimetric signals are sensitive to high-magnification
microlensing events or extended sources where we have the condition
of $u_{cm} \simeq \rho_{\star}$.

\section{Binary microlenses}
\label{twolens} In this section we investigate a possible relation
between the polarization and the astrometric centroid shift in
binary microlensing events. In these events, the polarization signal
when the source star crosses the caustic curve is higher than that
in the single lens case \cite{Agol96}. The caustic lines can be
classified into two categories of (i) folds where the caustic lines
are smooth and (ii) cusps where folds cross at a point
\cite{PettersW1996}. We will investigate the polarization and
astrometric signals near fold and cusp singularities.

Assuming that a source star crosses the fold of a caustic curve, two
temporary images can appear on the lens plane with almost the same
magnification factors \cite{Schneider92f}. On the other hand, there
are global images which do not change during caustic crossing. These
global images move slowly and have small magnification factors with
respect to the temporary images. Hence, we only consider the
temporary images and obtain their light centroid and polarization
vectors in our calculation. We set the centre of coordinate located
at the fold and axes are parallel with and normal to the tangent
vector of the fold. We also connect the source centre to the centre
of coordinate i.e. $\boldsymbol{u}_{cm}=(0,z\rho_{\star})$,
therefore, the position of any point on the source surface in this
coordinate is given by
$\boldsymbol{u}=\boldsymbol{u}_{cm}+\rho_{\star}\boldsymbol{\rho}$.
Since the position vector of local images due to the fold
singularity, $\boldsymbol{\theta}_{\pm}$
%given by equation (\ref{position}),
is a linear function with respect to
$\boldsymbol{u}$, hence
$\boldsymbol{\theta}_{\pm}(\boldsymbol{u})=\boldsymbol{\theta}_{\pm}(\boldsymbol{u}_{cm})
+\rho_{\star}\boldsymbol{\theta}_{\pm}(\boldsymbol{\rho})$. The
magnification factor for these images
%(equation \ref{magni2})
is
$$\mu_{\pm}=\frac{1}{2}\sqrt{\frac{u_{f}}{\rho_{\star}(\rho_{2}+z)}}$$
where $\rho_{2}\geq -z$. The light centroid of images near the fold
singularity is given by \cite{gaudi2001}:
\begin{eqnarray}
\boldsymbol{\theta}^{fs}_{f}&=&\boldsymbol{\theta}_{f,cm}+ \frac{
\sqrt{u_{f}\rho_{\star}}}{\pi\mu^{fs}_{f}ad}\int_{max(-z,-1)}^{1}
d\rho_{2} \int_{-\sqrt{1-\rho^{2}_{2}}}^{\sqrt{1-\rho^{2}_{2}}}\nonumber\\
&&d\rho_{1}I(\rho) \frac{1}{\sqrt{\rho_{2}+z}} \left( \begin{array}{c} d\rho_{1}-b\rho_{2}\\
-b\rho_{1}\end{array}\right),
\end{eqnarray}
where $\rho=\sqrt{\rho^{2}_{1}+\rho^{2}_{2}}$ and
$\boldsymbol{\theta}_{f,cm}=(\frac{-b\rho_{\star}}{ad}z,0)$ which is
parallel with the $u_{1}$-axis. For a limb-darkened source star, $y$
component of the second sentence vanishes due to the symmetric
range of $\rho_{1}$. Hence, near the fold singularities the centroid
shift vector is parallel with the caustic (i.e. $u_{1}$).

\begin{figure}
\begin{center}
\psfig{file=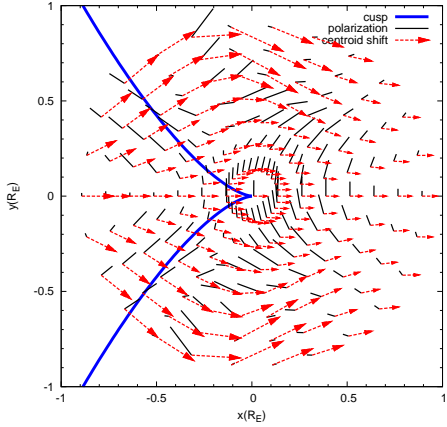,angle=270,width=8.cm,clip=} \caption{The
polarization (black lines) and astrometric centroid shift (red
vectors) near the cusp in a binary lens (blue line). We set
$\rho_{\star}=0.5$. Note that, the size of centroid shifts and
polarizations are divided by constant factors. Whenever the limb of
the source star crosses the cusp, the polarization
signal strongly changes.
} \label{cuspfig}
\end{center}
\end{figure}

On the other hand the Stokes parameters near the fold are given by:
 \begin{eqnarray}
\left( \begin{array}{c}
S_{Q}\\S_{U}\end{array}\right)&=&\sqrt{\frac{u_{f}}{\rho_{\star}}}\int_{max(-z,-1)}^1
d\rho_{2}\int_{-\sqrt{1-\rho^{2}_{2}}}^{\sqrt{1-\rho^{2}_{2}}}\nonumber\\&&d\rho_{1}
I_{-}(\rho) \frac{1}{\rho^{2}\sqrt{\rho_{2}+z}}\left( \begin{array}{c} \rho^{2}_{1}-\rho^{2}_{2} \\
2\rho_{1}\rho_{2} \end{array} \right).
\end{eqnarray}
Here, the second component (i.e. $S_{U}$) due to the symmetry of
integral over $\rho_{1}$ is zero. As a result, near the fold
singularities polarizations are normal to the tangent vector to the
fold and the centroid shift vectors. According to this orthogonality
relation near the fold singularity, the polarimetry and astrometry
can provide the following information of (i) tracing the source
trajectory with respect to caustic lines, (ii) in the case that the
orthogonality relationship is broken, we can study possible effects
of anomalies such as spots over the source surface. The
orthonormality relation can also be broken in cusp singularities
except for the case that the source is located at symmetric axis of
the cusp. For other points around the cusp, this orthogonal relation
does not exit. We calculate numerically the polarization and
centroid shift vectors around the cusp shown in Figure
(\ref{cuspfig}).

%%%%%%%%%%%%%%%%%%%%%%%%%%%%%%%%%%%%%%%%%%%%%%%%%%%%%%%%%%%%%%%%%%%%%%%%%%%%%%%%
\section{The effect of source spots on polarization and centroid shift}\label{spot}
In this section we study the effect of source spots on polarization
and centroid shift of microlensing events. Our aim is to extract the
astrophysical information of the spots and the atmosphere of the
source star from these anomalies.

A given spot on the source star with a different temperature and
magnetic field in comparison to the background star produces an
angular-dependent defect on the source surface. The result is
producing a net polarization even in the absence of the
gravitational lensing. The gravitational lensing can amplify the
polarization as well as change the orientation of the polarization.
The amount of magnified polarization depends on relative position of
spots with respect to the source centre and the lens position. Also
it depends on the temperature and intrinsic flux of spots.

Here we use three parameters to quantify a spot in our calculation
as: (i) the size of the spot, (ii) the temperature of the source
star and its temperature difference with respect to the spot and
(iii) the location of the spot on the source surface. Let us
characterize the lens plane by $(u_{1},u_{2})$ axes and put the
centre of lens at the centre of coordinate system. The projected
positions of the source and the spot in this reference frame are
$(u_{1,\star},u_{2,\star})$ and $(u_{1,s},u_{2,s})$. The radius of
the spot is $r_{s}$ and its angular size in the coordinate system
located at the centre of source star is given by
$\theta_{0}=\sin^{-1} (r_{s}/ R_{\star})$. For simplicity we choose
a circular spot over the source surface. In order to locate a
typical spot on the source star, we first put the position of the
spot at the star pole and then perform a coordinate transformation
and move the spot to an arbitrary location of the source
\cite{Mehrabi2013}. The position of the spot located at the pole of
the spherical coordinate system is given by:
\begin{eqnarray}
X_{s}&=&R_{\star} \sin\eta \cos\varphi\nonumber\\
Y_{s}&=&R_{\star} \sin\eta \sin\varphi\nonumber\\
Z_{s}&=&R_{\star} \cos\eta,
\end{eqnarray}
where $\eta$ and $\varphi$ change in the ranges of $[0,\theta_{0}]$
and $[0,2 \pi]$, respectively. The spot position projected on the lens
plane and normalized to the Einstein radius is
\begin{eqnarray}
x_{s}&=&\rho_{\star} \sin\eta \cos\varphi\nonumber\\
y_{s}&=&\rho_{\star} \sin\eta \sin\varphi\nonumber\\
z_{s}&=&\rho_{\star} \cos\eta,
\end{eqnarray}
and finally the spot position on the
lens plane, using two rotation angles
of $\theta$ around $y$-axis and $\phi$ around
$z$-axis is given by
\begin{eqnarray}
u_{1,s}&=&x_{s} \cos\phi \cos\theta  -y_{s} \sin\phi +z_{s} \cos \phi\sin\theta  + u_{1,\star}\nonumber\\
u_{2,s}&=&x_{s} \sin\phi \cos\theta  +y_{s} \cos\phi + z_{s} \sin\phi\sin\theta  + u_{2,\star}.
\end{eqnarray}
The modified Stokes parameters $S'_{Q}$, $S'_{U}$ and $S'_{I}$ for
the case of a single spot on the source are given by:
\begin{eqnarray}\label{spott}
\left(\begin{array}{c}S'_{Q}\\S'_{U}\end{array}\right)&=&
\left(\begin{array}{c}S_{Q}\\S_{U}\end{array}\right)-f\int_{\mathcal{A}_s}d^{2}s~I_{-}(\rho)A(u_{s})\left(\begin{array}{c}-\cos2\phi
\\\sin 2\phi\end{array}\right)\nonumber\\
&=&\left(\begin{array}{c}S_{Q}\\S_{U}\end{array}\right)-f
\left(\begin{array}{c}S_{Q,s}\\S_{U,s}\end{array}\right),\nonumber\\
S'_{I}&=&S_{I}-f\int_{\mathcal{A}_s}d^{2}s~I(\rho)A(u_{s})=S_{I}-fS_{I,s},
\end{eqnarray}
where $S_{Q}$, $S_{U}$ and $S_{I}$ are given by the equation
(\ref{tsparam}), these parameters assigned for the source star
without any spot and ${\mathcal{A}_s}$ represents area that is
covered by the spot. $u_{s}=\sqrt{u_{1,s}^2+u_{2,s}^2}$ is the
distance of each point of the spot from the lens position and
$f=\left[F_{\star}(\nu)-F_{\mathcal{A}_s}(\nu)\right]/F_{\star}(\nu)$
is the relative difference in the flux of the source and spot at the
frequency of $\nu$. We assume a black-body radiation for both star
and spot. The polarization degree $P'$ also can be given by:
\begin{eqnarray}\label{polt}
P'=\frac{\left(P^2 S_{I}^{2}+ f^2
P_{s}^{2}S_{I,s}^{2}-2fC(P,P_{s})S_{I}S_{I,s}\right)^{1/2}}{S_{I}-fS_{I,s}},
\end{eqnarray}
where we define $P_{s}\equiv({S_{Q,s}^2+S_{U,s}^2})^{1/2}/S_{I,s}$
and $C(P,S_{s})\equiv(S_{Q}S_{Q,s}+S_{U}S_{U,s})/S_{I}S_{I,s}$ as
the cross term between the contribution from the star and spot. The
angle of polarization $\theta'_{p}$ in terms of the Stocks
parameters of the source and spot is given by
\begin{eqnarray}\label{tetap}
\theta'_{p}&=&\frac{1}{2}\tan^{-1}[\frac{S_{U}}{S_{Q}}+f\frac{S_{U}S_{Q,s}-S_{Q}S_{U,s}}{S^{2}_{Q}}\nonumber\\
&+&f^2\frac{-S_{Q,s}S_{U,s}S_{Q}+S_{U}S^{2}_{Q,s}}{S^{3}_{Q}}+ ... ].
\end{eqnarray}
We can rewrite this equation in terms of unperturbed angle $
\theta_p$ and first- and second-order terms in $f$ as follows:
\begin{eqnarray}\label{tetap2}
\theta'_{p} &=& \theta_{p}+f\frac{S_{U}S_{Q,s}-S_{Q}S_{U,s}}{2P^{2}S^{2}_{I}}\nonumber\\&+&f^{2}\frac{C(P,P_{s})S_{I,s}}{2P^{4}S^{3}_{I}}(S_{U}S_{Q,s}-S_{Q}S_{U,s}) + ...
\end{eqnarray}

\begin{figure}
\begin{center}
\psfig{file=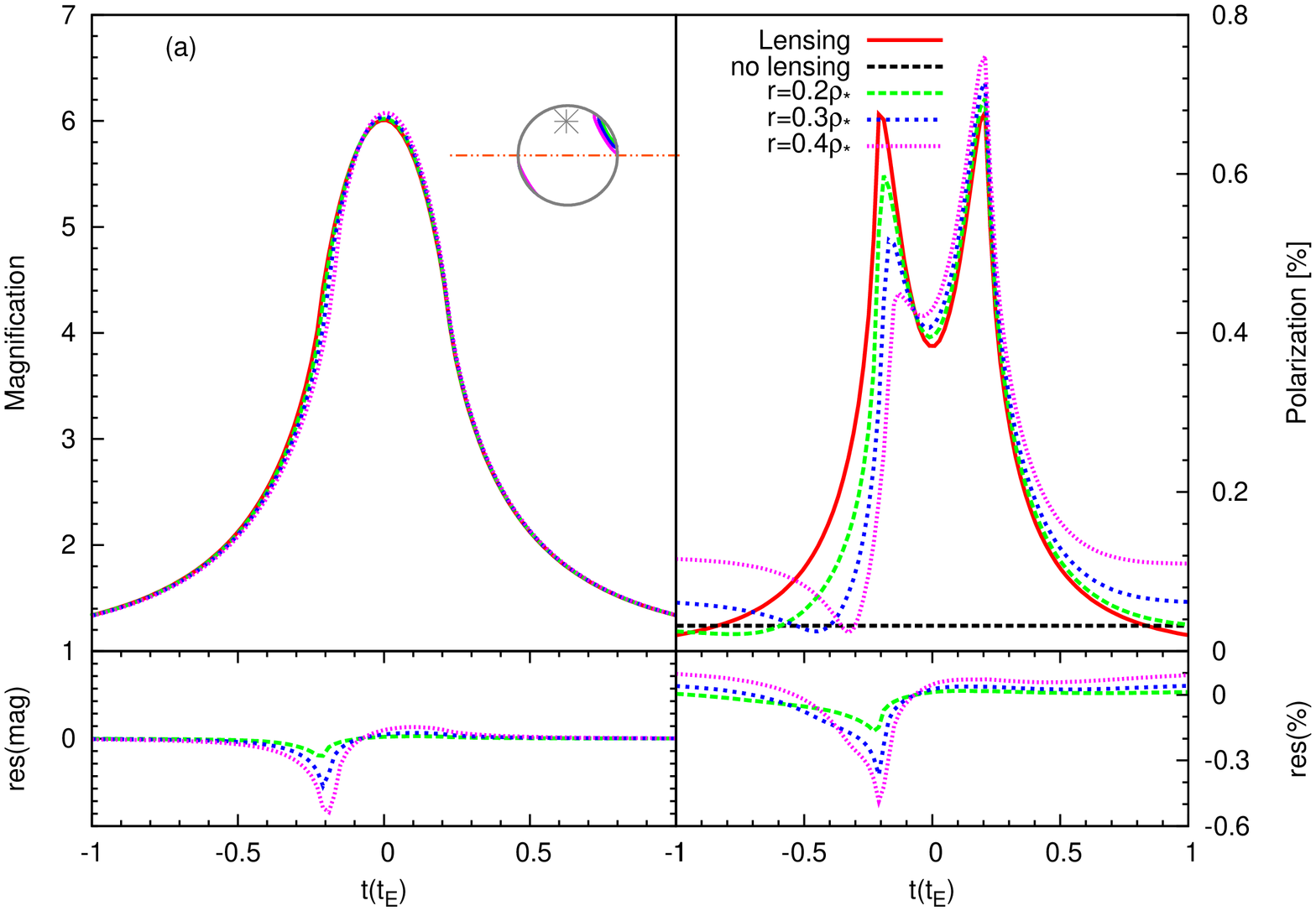,angle=270,width=8.cm,clip=}
\psfig{file=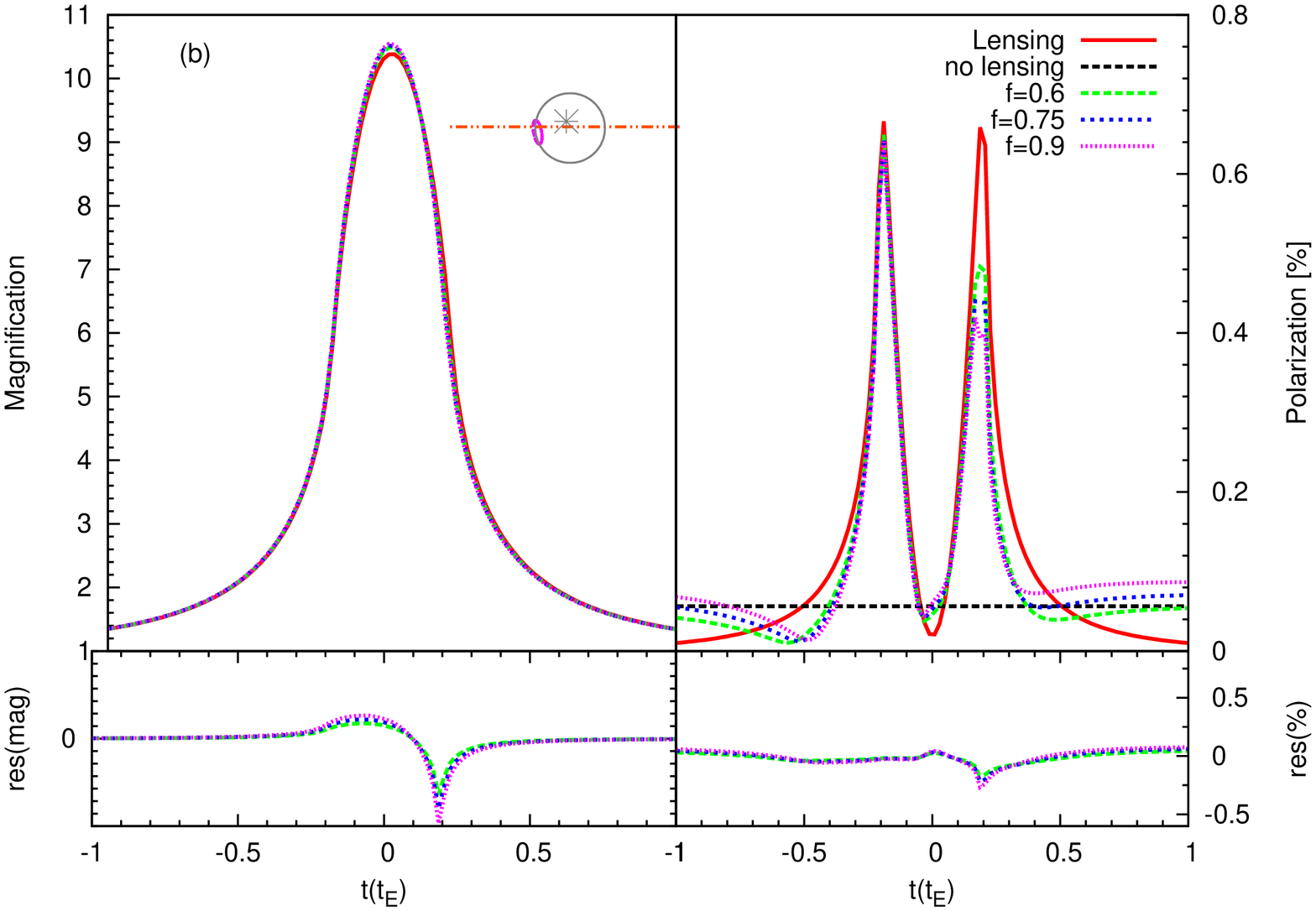,angle=270,width=8.cm,clip=}\caption{The two
microlensing events with spot on sources. In both subfigures (a) and
(b), the light curves and polarimetric curves with different set of
parameters for the spot are shown in left and right panels, respectively. The
source (grey circle) and its spot (pink spot), lens position (grey
star) and source centre trajectory projected in the lens plane (red
dash-dotted line) are shown with insets in the left-hand panels. The
simple models without spot effect are shown by red solid lines. The
black horizontal dashed lines represent the polarization signals due
to spotted sources without lensing effect. The photometric and
polarimetric residuals with respect to the simple models are plotted
in bottom panels. The parameters of these microlensing events shown
in subfigure $(a)$ and $(b)$ are $\rho_{\star}=0.3$,
$\theta=80^{\circ}$, $\phi=30^{\circ}$, $u_{0}=-0.205$, $f=0.9$ and
$\rho_{\star}=0.21$, $\theta=110^{\circ}$, $\phi=10^{\circ}$,
$u_{0}=-0.033$, $r_{s}=0.35\rho_{\star}$, respectively.}
\label{fig6}
\end{center}
\end{figure}
In Figures (\ref{fig6}) we represent two microlensing events
considering a stellar spot on the source surface. In both cases, we
plot the photometric and polarimetric light curves. The parameters
of these microlensing events shown in subfigures $(a)$ and $(b)$ are
$(a)$ $\rho_{\star}=0.3$, $\theta=80^{\circ}$, $\phi=30^{\circ}$,
$u_{0}=-0.205$, $f=0.9$ and $(b)$ $\rho_{\star}=0.21$,
$\theta=110^{\circ}$, $\phi=10^{\circ}$, $u_{0}=-0.033$,
$r_{s}=0.35\rho_{\star}$, respectively. Interpreting Figure
(\ref{fig6}), the polarimetric curve in the absence of stellar spots
for the case of $u_{0}<\rho_{\star}$ has two symmetric peaks at
$t(t_{E})=t_{0}\pm \sqrt{\rho_{\star}^{2}-u_{0}^2}$ where $t_{0}$ is
the time of the closest approach. For the case of a spot on the
source surface, not only the symmetry in the pholarimetic curve
breaks, also it tends to non-zero polarization at no-lensing stages.

\begin{figure}
\begin{center}
\psfig{file=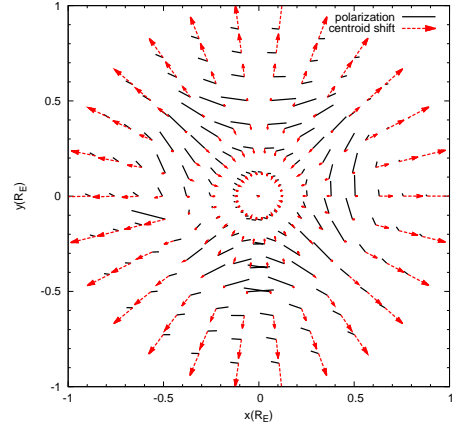,angle=270,width=8.cm,clip=} \caption{The
normalized polarization (black lines) and astrometric centroid shift
(red vectors) for a source star with stellar spot lensed by a point-mass
lens with the parameters of
$\rho_{\star}=0.5$, $\theta_{0}=18^{\circ}$, $\theta=110^{\circ}$,
$\phi=15^{\circ}$ and $f=0.75$. Note that, the size of centroid
shifts and polarizations are normalized by an arbitrary factor.}
\label{spotf}
\end{center}
\end{figure}

A significant signal of the stellar spot in microlensing events
happens when the lens approaches close enough to the spot. In that
case the total flux due to the spot $S_{I,s}$ increases and as a
result denominator of equation (\ref{polt}) decreases. On the other
hand, the cross term $C(P,P_{s})$ enhances, as $P$ is approximately
parallel with $P_{s}$ (noting that, the polarimetric vectors over
the spot are similar to the polarimetric vectors over the source
except having different temperatures). Moreover, if the spot is
located on the limb of the source star, one of peaks in the
polarimetric curve will strongly be disturbed. According to Figure
(\ref{fig6}), the larger and darker spots make stronger polarimetric
and photometric signals. We note that the presence of the spot has
small effects on light curves.

The astrometric centroid shift of a source with a spot is given by:
\begin{eqnarray}
\boldsymbol{\delta\theta}'_{c}=\frac{1}{S'_{I}}\{ S_{I}
\boldsymbol{\theta}_{c} -f\int_{\mathcal{A}}d^{2}s~ I(\rho)
\frac{u_{s}^2+3}{u_{s}\sqrt{u_{s}^2+4}}\nonumber\\
\left( \begin{array}{c} \rho \cos\phi+ u_{cm} \\
\rho\sin\phi\end{array} \right)\}-\boldsymbol{u}_{spot},
\end{eqnarray}
where $\boldsymbol{\theta}_{c}$ is the light centroid vector for a
source star without any spot (see equation \ref{astrop}) and
$\boldsymbol{u}_{spot}$ is the light centroid of the source with a
spot. The spot perturbation on the astrometric measurements is
larger for the smaller impact parameters. In Figure (\ref{spotf}),
we plot the map of normalized polarization and centroid shift
vectors of the source with a spot and the parameters of
$\rho_{\star}=0.5$, $r_{s}=0.3\rho_{\star}$, $\theta=110^{\circ}$,
$\phi=15^{\circ}$, and $f=0.75$, lensed by a single lens. We note
that the orientations of polarization and centroid shift in the
absence of the spot are orthogonal. The polarization vector changes
strongly due to the spot while the centroid shift is almost
unchanged except for the close approaches. As a result, the
anomalies over the source surface break the orthogonality relation
between the astrometric centroid shift and polarization vectors.

\subsection{Statistical investigation of the OGLE data for polarimetry observation}
Here we investigate the statistics of high magnification events in
the list of the OGLE data to identify the fraction of events with
possible signatures of the polarization. For the case of high
magnification events with $u_{0}<\rho_{\star}$, the time profile of
polarization has two peaks and the maximum polarization may reach to
one percent. The large telescopes with enough exposure times can
achieve the sensitivity of detecting the polarization as well as
time variation of this parameter. In a recent work by Ingrosso et
al. (2012), they prospect the observation of the polarization by
FOcal Reducer and low dispersion Spectrograph 2 (FORS2) on the Very
Large Telescope (VLT). A detailed study for follow-up observations
of polarization of the OGLE microlensing data will be presented in a
later work.

\begin{figure}
\begin{center}
\psfig{file=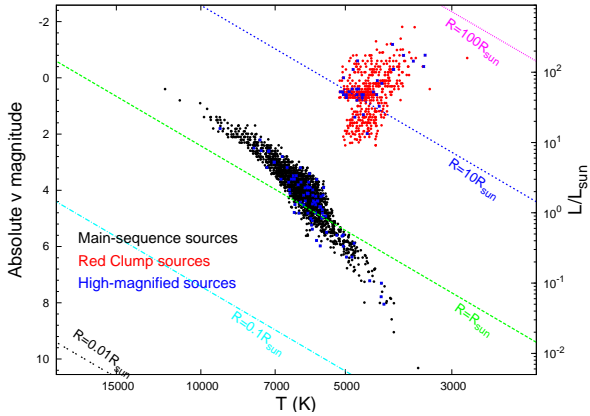,angle=270,width=8.cm,clip=}
\caption{Temperature-Luminosity diagram for the OGLE microlensing events
with main-sequence (black points) and red clump sources (red
points). The blue stars represent high magnified sources with
$u_{0}<\rho_{\star}$. The lines in this diagram show the constant
radii for the stars.} \label{colorM}
\end{center}
\end{figure}

Here, we examine $3560$ microlensing events reported by the OGLE
collaboration towards the Galactic bulge for the period of years
2001-2009 \cite{OGLE3}. These microlensing events have been detected
by monitoring $150$ million objects in the Galactic Bulge. Amongst
microlensing events listed by the OGLE collaboration, for $2614$
number of events, the position of source stars are identified in the
Colour Magnitude (CM) diagram. The source stars in CM are grouped in
(i) the main-sequence stars (assigned by black dots) and (ii) red
clump stars (assigned by red dots) in Figure (\ref{colorM}).

In order to estimate the relevant parameter in the polarization
(i.e. $u_0/\rho_\star$), we use the best value of $u_0$ from fitting
to the observed light curve for single lensing. On the other hand we
estimate the size of source star from radius-CM diagram, noting that
the absolute magnitude verse colour of starts are corrected by the
amount of the extinction and the reddening estimated from the
Besan\c{c}on model in the Galaxy \cite{Robin2003}. For calculating
the radius of star, we use the Stefan--Boltzmann law (i.e. $L=4\pi
R^2\sigma T^{4}$) and for a given luminosity and temperature, we
obtain the radius of star. The straight lines in Figure
(\ref{colorM}) represent the constant radius lines for the stars in
CM diagram. Now we need the calculation of $\rho_{\star}$, we set
the lens mass $M_L = 0.3 M_{\odot}$, distance of source $D_{s}=8.5$
kpc and distance of lens at $x=0.5$ where the probability of lensing
is maximum.

\begin{figure}
\begin{center}
\psfig{file=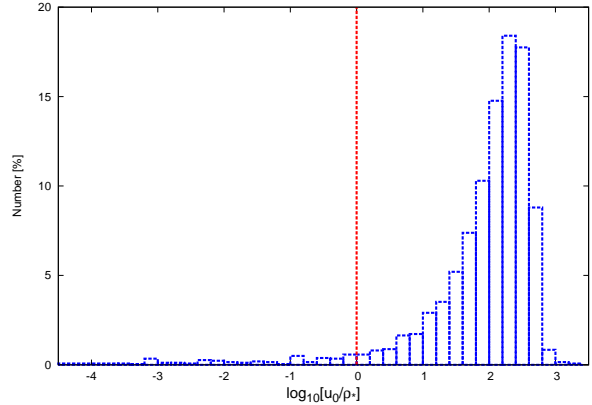,angle=270,width=8.cm,clip=} \caption{The
distribution of $u_{0}/\rho_{\star}$ for the OGLE-III microlensing
events. Almost $4.3$ percent of these events are in the range of
$u_{0}<\rho_{\star}$. These events are potentially suitable targets
for measuring the polarization due to spots on the source
star.} \label{urho}
\end{center}
\end{figure}

The distribution of $u_{0}/\rho_{\star}$ for the OGLE microlensing
events is shown in Figure (\ref{urho}). Amongst $2614$ number of
microlensing events, the high magnification events with the
criterion of $u_{0}<\rho_{\star}$ contains (a) $81$ source stars in
the main sequences and (b) $32$ source stars in the red clumps.
Using this statistics, we estimate that almost $4.3$ per cent of
microlensing events satisfy the condition of $u_0/\rho_\star<1$. For
these events the polarization due to possible spots on the source
surface with a suitable device is observable.

\section{conclusions}
\label{result} In gravitational microlensing observations, the
tradition is the follow-up photometry of ongoing events. This
strategy of observation is aimed to produce precise light curves
from microlensing events with high cadences and small photometric
error bars. We can imagine performing two more types of astrometric
and polarimetric observations of microlensing events. These
observations are aimed to measure the time variations of centroid
shift of images and polarization during the lensing.

We took a perfect circular shape as the source star for the
microlensing. In gravitational microlensing the symmetry of source
star is broken by producing two distorted images at either side of
the lens. This breaking of symmetry causes some phenomena such as
the polarization and light centroid shift of source star images. We
have showed that there is an orthogonality relation between the
polarization and light centroid shift vectors in simple microlensing
events. The observations of either polarimetry or astrometry can
uniquely indicate the source trajectory on the lens plane. By
exploring the sign of impact parameter and source trajectory, all
related degeneracies i.e. $u_{0}$ degeneracy, ecliptic degeneracy,
parallax-jerk degeneracy and orbiting binary ecliptic degeneracy can
be resolved \cite{Skowron2011,rah11}. We noted that while the
polarimetry can probe the small impact parameters, the astrometry is
sensitive to the large impact parameters and applying these two
observations can probe all the ranges of impact parameter.

In the binary microlensing events, unlike to the simple lensing, the
orthogonality between the polarization and centroid shift generally
is not valid. During the caustic crossing of the source star which
produces the maximum signals of polarization and centroid shift, we
studied the behavior of these parameters for the fold and cusp
singularities. We have shown that orthogonality relation between the
polarization and centroid shift is valid for the fold singularity
while this relation is violated in the cusp singularity. This
behavior between these two vectors during microlensing events can be
an indicator to identify the trajectory of the source with respect
to the caustic lines in the binary lensing.

Finally, we have studied the effect of the source spots on
polarization, astrometry and photometry of microlensing events for
the single lensing. One of features of spots is breaking the
orthogonality relation between the polarization and centroid shift.
Studying the map of polarization and centroid shift with the
photometry provides some information on the physics of the spots on
a source star. We have investigated the high magnification events in
the list of the OGLE data \cite{OGLE3} and show that for $4.3$
percent of events the polarization effect could be enhanced with the
amount of about $1$ percent. This means that for a large number of
very high magnification microlensing events, the polarimetry
follow-up observations can open a new window for studying the
stellar spots on various types of stars.

\textbf{Acknowledgment} We are grateful to Philippe Jetzer and Reza
Rahimitabar for reading and commenting on the manuscript.
%%%%%%%%%%%%%%%%%%%%%%%%%%%%%%%%%%%%%%%%%%%%%%%%%%%%%%%%%%%%%%%%%%%%%%%%%%%%%


\begin{thebibliography}{}

\bibitem[Akitaya et al. 2009]{Akitaya2009}%********************************************
Akitaya H., Ikeda Y.,  Kawabata K.s., et al., \ 2009, A \& A, 499,
L163.

\bibitem[Agol 1996]{Agol96}%*************************************************8
Agol E., \ 1996, MNRAS, 279, L571.

\bibitem[Blnadford \& Narayan 1986]{Blnadford}%***********************************
Blnadford R.D., \& Narayan R., \ 1986, ApJ, 310, L568.


\bibitem[Bogdanov et al. 1996]{Bogdanov96}%********************************************
Bogdanov M. B., Cherepashchuk A. M. \& Sazhin M. V., \ 1996, Ap \&
SS, 235, L219.

\bibitem[Chandrasekhar 1960]{chandrasekhar60}%********************************************
Chandrasekhar S., \ 1960, Radiative Transfer. Dover Publications,
New York.

\bibitem[Chang \& Refsdal 1979]{ChangRefesdal}%***********************************
Chang K., Refsdal S., \ 1979, Nature, 282, L561.

\bibitem[Chung et al. 2009]{Chung}%**********************************
Chung S.-J., Park B.-G., Ryu Y.-H. \& Humphrey A. \ 2009, APJ, 695,
L1357.

\bibitem[Drissen et al. 1989]{Drissen1989}%******************************************8
Drissen L., Bastien P., \& St.-Louis N., \ 1989, ApJ, 97, L814.


\bibitem[Dominik 1999]{Dominik}%************************************
Dominik M. \ 1999, ApJ, 522, L1011.

\bibitem[Dominik 2004]{Dominik2004}%*******************************************
Dominik M., \ 2004, MNRAS, 353, L69.

\bibitem[Einstein 1911]{Einstein11}%*************************************************
Einstein A. \ 1911, Annalen der physik, 35, L898.

\bibitem[Einstein 1936]{Einstein36} %******************************************
Einstein A., \ 1936, Science, 84, L506.

\bibitem[Gaudi 2012]{gaudi2012}%********************************************
Gaudi B.s., \ 2012, A. R. A\& A, 50, L411.

\bibitem[Gaudi \& Petters 2002]{gaudi2001}%%%%%%%%%%%% fold*************************
Gaudi B.S. \& Petters A.O., \ 2002, ApJ, 574, L970.

\bibitem[Gaudi \& Petters 2002]{gaudi2002}%%%%%%%%%%%%%%% cusp***************************
Gaudi B.S. \& Petters A.O., \ 2002, ApJ, 580, L468.

\bibitem[Gould \& Han 2000]{Gould}%***************************************
Gould A. \& Han C. \ 2000, ApJ, 538, L653.

\bibitem[H{\o}g et al. 1995]{Hog}% astro....********************************
H{\o}g, E., Novikov, I. D., \& Polnarev, A. G. \ 1995, A \& A , 294,
L287.

\bibitem[Hayashi et al. 1962]{Hayashi62}
Hayashi C., H\={o}shi R., Sugimoto D., \ 1962, Prog. Theor. Phys.
Suppl., 22, L1.

\bibitem[Ingrosso et al. 2012]{Ingrosso}%******************************************
Ingrosso, G., Calchi Novati,S., De Paolis F., et al. \ 2012, MNRAS,
426, L1496.

\bibitem[Jeong et al. 1999]{Jeong}%************************************
Jeong Y., Han C. \& Park S.-H. \ 1999, ApJ,511, L569.

\bibitem[Liebes 1964]{Leibes}%*****************************************
Liebes, Jr.S., \ 1964,  Phys. Rev., 133, L835.

\bibitem[Mehrabi \& Rahvar 2014]{Mehrabi2013}%***********************************
Mehrabi A. \& Rahvar S., \ 2014, In preparation.

\bibitem[Miralda-Escd\'e 1996]{Miralda96}%***********************************
Miralda-Escud\'e J., \ 1996, ApJ, 470, L113.

\bibitem[Miyamoto \& Yoshii 1995]{Miyamoto}% astro....***************************
Miyamoto, M. \& Yoshii, Y. \ 1995, AJ, 110, L1427.

\bibitem[Mao \& Witt 1998]{Witt98}%******************************888
Mao S., \& Witt H.J., \ 1998, MNRAS, 300, L1041.

\bibitem[Paczy\'nski 1986a,b]{Paczynski86}%***********************************
Paczy\'nski B., \ 1986a, ApJ, 301, L502.

\bibitem[]{paczynski86b}%*******************************************
Paczy\'nski B., \ 1986b, ApJ, 304, L1.

%\bibitem[Paczy\'nski 1970]{Paczynski70} Paczy\'nski B., \ 1970,
%Acta Astronomica, 20, L47.

\bibitem[Paczy\'nski 1997]{Paczynski96}%*********************************
Paczy\'nski B., \ 1997, Astrophys. J. Lett. astro-ph/9708155.

\bibitem[Petters \& Witt 1996]{PettersW1996}%***********************************
Petters A. O., \& Witt H. J., \ 1996, J. Math. Phys., 37 L2920.

\bibitem[Rahvar \& Ghassemi 2005]{Rahvar2005}%********************************
Rahvar, S.,  \& Ghassemi, S. \ 2005, A \& A, 438, L153.

\bibitem[Rahvar \& Dominik 2009 ]{rah11} Rahvar, S., \& Dominik, M.\ 2009, MNRAS, 392, 1193

%\bibitem[Refsdal \& Weigert 1971]{Refesdal71} Refesdal S., \&
%Weigert A., \ 1971, A \& A, 6 L426.

\bibitem[Robin et al. 2003]{Robin2003}
Robin, A. C., et al., \ 2003, A \& A, 409, L523.

\bibitem[Schneider \& Wagoner 1987]{schneider87}%*****************************************
Schneider P., Wagoner R. V., \ 1987, ApJ, 314, L154.

\bibitem[Schneider 1985]{Schneider85}%*****************************************
Schneider P., \ 1985,  A \& A, 143, L413.

\bibitem[Schneider et al. 1992a]{Schneider92f}%**************************************88
Schneider P., Ehlers J., \& Falco E.E \ 1992a, Gravitational lenses,
Berlin: Springer Verlag.

\bibitem[Schneider \& Weiss 1992b]{Schneider92c}%**************************************
Schneider P., \& Weiss A., \ 1992b, A \& A, 260, L1.

%\bibitem[]{Schonberg42}
%Sch\"{o}nberg, M., \&  Chandrasekhar S., \ 1942,  ApJ, 96, L161.

\bibitem[Simmons et al. 1995a,b]{simmons95a}%**********************************************
Simmons J. F. L., Newsam A. M. \& Willis J. P., 1995a, MNRAS, 276,
L182.

\bibitem[Simmons et al. 1995b]{simmonsb}%***************************************
Simmons J. F. L., Willis J. P. \& Newsam A. M., 1995b, A \& A, 293,
L46.

\bibitem[ Sobolev 1975]{sob}
Sobolev, V.V. (1975) Light Scattering in Planetary Atmospheres, Pergamon Press, Oxford.

\bibitem[Skowron et al. 2011]{Skowron2011}%***********************************************
Skowron J. et al., \ 2011, ApJ, 738, L87.

\bibitem[Tinbergen 1996]{Tinbergen96}%*******************************************
Tinbergen J., \ 1996, Astronomical Polarimetry. Cambridge Univ.
Press, New York.


\bibitem[Walker 1995]{Walker}%*****************************
Walker M.A. \ 1995, ApJ, 453, L37.

\bibitem[Wyrzykowski et al. 2014]{OGLE3}
Wyrzykowski, {\L}., et al., 2014, arXiv:1405.3134.


\bibitem[Yoshida 2006]{yoshida06}%***********************************
Yoshida H., \ 2006, MNRAS, 369, L997.
\end {thebibliography}
\end{document}